\documentclass{interact}
\pdfoutput=1
\usepackage[utf8]{inputenc}
\usepackage[english]{babel}
\usepackage{hyperref}
\usepackage{amsmath}
\usepackage{csquotes}
\usepackage{xurl}
\usepackage{lmodern}
\usepackage[notes,backend=biber]{biblatex-chicago}
\bibliography{main}

\begin{document}

\title{Synthesizing Proteins on the Graphics Card. Protein Folding and the Limits of Critical AI Studies}

\author{Fabian Offert, Paul Kim, Qiaoyu Cai}

\maketitle

\begin{abstract}
This paper investigates the application of the transformer architecture in protein folding, as exemplified by DeepMind's AlphaFold project, and its implications for the understanding of so-called large language models. The prevailing discourse often assumes a ready-made analogy between proteins -- encoded as sequences of amino acids -- and natural language, which we term the language paradigm of computational (structural) biology. Instead of assuming this analogy as given, we critically evaluate it to assess the kind of knowledge-making afforded by the transformer architecture. We first trace the analogy's emergence and historical development, carving out the influence of structural linguistics on structural biology beginning in the mid-20th century. We then examine three often overlooked preprocessing steps essential to the transformer architecture, including subword tokenization, word embedding, and positional encoding, to demonstrate its regime of representation based on continuous, high-dimensional vector spaces, which departs from the discrete nature of language. The successful deployment of transformers in protein folding, we argue, discloses what we consider a non-linguistic approach to token processing intrinsic to the architecture. We contend that through this non-linguistic processing, the transformer architecture carves out unique epistemological territory and produces a new class of knowledge, distinct from established domains. We contend that our search for intelligent machines has to begin with the \textit{shape}, rather than the \textit{place}, of intelligence. Consequently, the emerging field of critical AI studies should take methodological inspiration from the history of science in its quest to conceptualize the contributions of artificial intelligence to knowledge-making, within and beyond the domain-specific sciences.

\end{abstract}

\section{Introduction}\label{introduction}

The use of large transformer-based machine learning models to predict the three-dimensional structure of proteins from amino acid sequences (``protein folding'') has been widely perceived as one of the most important achievements in contemporary artificial intelligence research. \autocites{moussad_2023,metz_2023,offert_2021,raghu_2020} AlphaFold, released by DeepMind in 2020, was the first model to suggest this view. ``It will change everything,'' a 2020 \textit{Nature} article quotes evolutionary biologist Andrei Lupas. \autocite{callaway_2020} ``This will change medicine. It will change research. It will change bioengineering.'' \autocite[1]{callaway_2020} And while the unfettered optimism of four years ago has since transformed into a more cautious assessment of AlphaFold's capabilities, \autocite{terwilliger_2023} protein folding -- and one of its main projected ``downstream'' applications, drug discovery -- is still cited as one of the most ``exceptionally useful'' \autocite{terwilliger_2023} use cases for large transformer-based machine learning models. Most recently, the utilization of a large transformer-based machine learning model to design new CRISPR candidate proteins \autocite{ruffolo_2024} -- ``blueprints for microscopic mechanisms that can edit your DNA'', as a \textit{New York Times} \autocite{metz_2024} report on the discovery reads -- has promised even more immediate returns. It is perhaps not surprising, then, that between the first and final submissions of this article, John Jumper and Demis Hassabis at Google received the 2024 Nobel Prize in chemistry for AlphaFold and its \textit{potential} impact on protein science.

At the same time, a different transformer-based model, the large language model ChatGPT, has dominated headlines ever since its release in 2022, eclipsing the comparably extensive coverage that AlphaFold, as a purely scientific model, and subsequent protein folding models had received by orders of magnitude (apart, maybe, from the reporting in the immediate aftermath of the Nobel Prize). ChatGPT's remarkable capability to both hold a casual conversation and produce working computer programs suggests an ``understanding'' of both natural and formal languages that goes far beyond that of previous generations of models, and points to a future in which computation itself becomes ``neural'' -- a future where even essential operations and calculations will be facilitated by a transformer layer. \autocites[See][]{offert_2023a,offert_2023b}

AlphaFold and ChatGPT, obviously, represent two very different use cases of transformers. But beyond their ``two cultures'' lies a strange commonality. Both systems' claim to fame is their ability to reveal the ``meaning''\footnote{Meaning is set in quotes here in reference to Hannes Bajohr's argument that, while what is learned by large language models appears meaning-like, it is missing many of the characteristics of what humans would designate as meaning. Bajohr suggest the term ``dumb meaning'' to describe this not-here-nor-there character of computational meaning-making. See \autocite{bajohr_2023}.} hidden in long sequences of tokens. ChatGPT is seen as a success because it can infer what is ``meant'' by a sequence of words; AlphaFold is seen as a success because it can infer what is ``meant'' -- i.e., which three-dimensional protein structure is described -- by a sequence of amino acids. This ability to infer ``meaning'' derives, technically, from the transformer's ability to model ``long-range relationships in the input sequence beyond their sequential neighborhoods.'' \autocite{moussad_2023} Both systems, in other words, due to their exposure to very large sets of training data, can ``see'' higher-order structure where humans cannot.

To the humanist, this development of course sounds like structuralism 2.0 \autocite{gastaldi_2024}, both due to its intuitive appeal -- we like to think that we are thinking in abstractions, so abstraction is thought of as thinking -- and due to the large number of potential applications. Structuralism's appeal was that of a meta-science -- much like that of cybernetics, to which it is historically and systematically related. \autocite{geoghegan_2023} The same meta-science appeal is reflected in statements like the following made by Ali Madani, whose company develops the protein folding model targeting CRISPR mentioned above: ``These A.I. models learn from sequences --- whether those are sequences of characters or words or computer code or amino acids.'' \autocite{metz_2024}

Its claim to universality, then, is the transformer's selling point. In the following, we would like to argue that this claim to universality, particularly when viewed historically, is misleading. Concretely, it masks the fact that the direction of the simile, overwhelmingly, points from language to proteins: ``Proteins, as a sequence of amino acids,'' -- as one recent paper reads -- ``can be viewed precisely as a language and therefore modeled using neural architectures developed for natural language.'' \autocite[1]{vig_2020} Another recent paper proposes that the ability of pre-trained language models ``to represent, manipulate, and extrapolate more abstract, nonlinguistic patterns may allow them to serve as basic versions of general pattern machines.'' \autocite{mirchandani_2023} In accounts like this, it is the language-ness of proteins (and of language), in other words, that renders the transformer architecture universal, and enables it to model two very different domains.

This expansion of the domain of language into structural biology seems ripe with implications for the emerging field \autocite{raley_2023} of critical AI studies which, so far, has focused almost exclusively on ``cultural'' models like ChatGPT. What is more, the critique of critical AI studies is almost always cultural critique in the cultural studies sense: a critique of culture, with culture. ``I know it when I see it,'' ``it'' being ``wrong'' culture, aesthetically, ethically, and ideologically. The model has produced a ``wrong'' picture, a ``wrong'' text, which points to the model's limitations. Such idiosyncratic methodologies \autocite{offert_2024} necessarily have to ignore scientific models which simply cannot be evaluated in this way.

In this paper, we suggest that, by all but ignoring the vast landscape of scientific models within and beyond structural biology, critical AI studies is missing an entire angle of critique that draws from the history of science in general, and the history of knowledge in particular. We ask: what kind of knowledge-making do transformers afford, independent of their application? And what can we learn, specifically, about the ``cultural'' model ChatGPT, from the ``scientific'' model AlphaFold and its many siblings and successors?

To answer these questions, in a first step, we trace the emergence and historical development of the language paradigm in biology which has guided epistemic assumptions about structural biological processes since the mid 20th century. Biological processes, at least since the discovery of the genetic code, have been understood as language-like, if not as ``linguistic,'' i.e.~as sharing structural properties (e.g.~a ``grammar'') with human language. The use of large transformer-based machine learning models can be understood as a direct continuation of this ``traditional'' language paradigm in computational biology in the 21st century.

In a second step, we argue that the transformer architecture works so well exactly because it is, counterintuitively, ``non-linguistic.'' While both natural language and protein folding models deal with sequences, this is where the commonality ends. Amino acids are not letters and amino acid sequences are not sentences.\footnote{They are, for instance, much, much longer.} Seemingly peripheral (as in, not directly connected to the attention mechanism) techniques like word embedding, positional encoding, and subword tokenization ``infuse'' tokens with a continuity that could not be more language-unlike, and actually is much closer, analogically, to a physical, not a symbolic understanding of the world. Word embedding, for instance, produces continuous concepts which are manifested as points in an embedding space and that are at the same time more and less ``ambiguous'' than words: they are precise as they are enumerable but at the same time they can not be reduced to any existing word, as they reside in the ``real'' (as in real numbers, not as in Lacan) gaps that emerge from modeling a set of discrete tokens as a continuous space. Positional encoding gives an additional ``spin'' to tokens that encodes their position in a sentence, a concept equally far removed from the discrete world of language. Subword tokenization, finally, dissolves the very concept of the word, and gives equal importance to partials and composites. Before any data can be attended to, in other words, it has already lost its linguistic character.

We finally posit, in a third step, that transformers have to be understood as approaches to knowledge-making distinct from domain-specific affordances. Not only are transformers not ``language models'' (an understanding that is already questioned by the mere existence of transformers in other cultural domains, for instance in the form of image models), they are also not ``linguistic'' in the sense that they share structural properties with human language. The intuitive (humanist, and implicitly structuralist) interpretation that the many capabilities of contemporary artificial intelligence systems are a result of the fact that ``all the world is language'' -- itself historically determined by the early Cartesian emphasis on linguistic compositionality \autocite{chomsky_2009} as a defining indicator of human-ness -- is thus a fallacy. Instead, we suggest that transformers allow for the operationalization of sequential intuition, a concept we model after Peter Galison's distinction between the ``image'' and ``logic'' modalities of scientific reasoning. Protein folding can be understood through sequential intuition. But as we are dealing with unknown (or at least under-formalized) bio-physical processes, their transformation into sequential operations is not straightforward and thus left to the machine. That machine, however, requires continuity (in the computational sense) to model continuity (in the physical sense) -- which means that any linguistic qualities introduced must immediately be dissolved again, through word embedding, positional encoding, and subword tokenization.

\section{The Problem of Protein Folding}\label{the-problem-of-protein-folding}

\subsection{Protein Folding with X-Ray Crystallography}\label{protein-folding-with-x-ray-crystallography}

Proteins are biomolecules that fulfill a variety of functions, including those that determine procreation (RNA polymerase). \autocite{gomes_2019} Their study is a comparably late addition to the biological sciences. Although their macroscopic analysis goes back to the late eighteenth and early nineteenth century it was only in 1949 that Frederick Sanger's sequencing of insulin showed that proteins indeed must consist of linear polymers of amino acids. Crucially, it is the three-dimensional structure of these polymers that determines a protein's biological function. But while the transcription and translation from DNA into amino acid sequences has been well studied, how exactly this three-dimensional structure, or native structure, is determined from an amino acid sequence, or primary structure, has long been an open problem, known as ``protein folding''. As a recent introduction to the subject puts it: ``Within this statement lies one of the mind-blowing facts in Protein Science---the realization that a given linear chain of amino acids encodes all the required information to fold the polypeptide into the native structure, as well as one of the major open questions in the field---what are the rules that dictate such specific and unique protein structure.'' \autocite[vii]{gomes_2019}

X-ray crystallography, a largely experimental method, at one point was the ``most favoured technique'' for determining structures of proteins and other biological macromolecules. \autocite[8]{smyth_2000} The diffracting nature of x-rays allows for the observation of diffraction patterns after exposing a crystal to x-ray beams. The diffraction patterns contain information on the crystal structure and the positions of atoms in the crystal.

Proteins, however, do not occur naturally as crystals. Though x-ray crystallography has been and remains an extremely useful method of experimentally determining three-dimensional protein structure, one of its major challenges has been the crystallization process itself. Multiple variables such as chemical precipitant, concentration of protein, and crystallization technique, among others, determine the feasibility of the crystallization process. As biochemists Michael Smyth and Jan Martin write: ``The growth of protein crystals of sufficient quality for structure determination is, without doubt, the rate limiting step in most protein crystallographic work, and is the least well understood.'' \autocite[8]{smyth_2000} Furthermore, not all proteins can be (easily) crystallized, and as such, not all protein structures can be determined by x-ray crystallization.~

While it has been continuously improved and even partially automated during more than a century of practice, x-ray crystallography thus remains an inefficient path towards large scale protein structure determination, simply due to the contingencies of its experimental nature. The promise of models like AlphaFold, then, is exactly to fulfill what Joseph \textcite{kraut_1965} hinted at already in 1965: ``Hopes have even been raised that it will someday be possible to deduce conformations solely from amino acid sequence.'' But, as he writes further: ``To be sure, there is reason to expect that before long the various kinds of interactions responsible for protein conformations will be correctly identified and measured. However, unless as-yet unsuspected rules of an improbably simple kind eventually emerge, it appears very unlikely that it will be possible to make sufficiently accurate calculations for them to be of much predictive value.''

\subsection{Protein Folding Turns Computational}\label{protein-folding-turns-computational}

In 1962, scientists Max Perutz and John Kendrew shared the Nobel Prize in Chemistry for successfully modeling the first complex protein, hemoglobin. Perutz and Kendrew used x-ray crystallography at increasingly high resolutions in order to determine this structure, resulting in tens of thousands of images to be analyzed per heavy atom derivative. \autocite{chadarevian_2018} Automation was a necessary component of this mapping endeavor. While Kendrew's usage of the EDSAC computer at the University of Cambridge never resulted in anything close to a full simulation of the protein folding process, it facilitated many arithmetic tasks that are necessary steps in the analysis process like calculating Fourier syntheses. \autocite{bennett_1952}

A decade later, computer simulation -- itself an emerging trading zone \autocite{galison_2011} -- promised to take automation a step further in the form of molecular dynamics. \autocites{levitt_1975,mccammon_1977} Molecular dynamics treats particles as simple objects that behave according to physical laws of motion within a computationally simulated environment. \autocite{mccammon_1977} This turn towards physics-based approaches came as a result of an increasing awareness of the complexity of the folding process. Instead of taking thousands of detailed images of three dimensional protein structures and analyzing them computationally like Perutz and Kendrew, molecular dynamics allowed for easy observations of particle interactions at the risk, of course, of drawing conclusions fully valid only within the simulated, simplified environment. Moreover, the problem of rate-limitation returned with a vengeance: to model the exact atomic environment of a protein was computationally expensive \autocite[1139]{hollingsworth_2018}, with a clear trade off between accuracy and efficiency.~

\subsection{Protein Folding and Computational Complexity}\label{protein-folding-and-computational-complexity}

By this time, Anfinsen's Dogma, which asserts that the primary structure of a small globular protein (the amino acid sequence) is the sole determinant of the native three-dimensional structure of the protein, had been established as a ground-level postulate. \autocite{anfinsen_1973} Levinthal's Paradox, a thought experiment that indicates that the high number of possible conformations of an amino acid sequence rules out a random folding search path, had been largely accepted by this point as well. \autocite{levinthal_1969} The paradox roughly goes as follows: If proteins achieved their final three-dimensional structures by iterating through every possible configuration, given that the number of configurations is so large, the time to fold would be almost innumerable scales of magnitude larger than the actual time it takes for a protein to fold.~

Levinthal's paradox on its own is worth a more thorough investigation than possible here. It should be pointed out, nevertheless, that it represents an ``application'' of computational complexity to a biological process, further underlining the entanglement of ``model and machine'' \autocite{dick_2015} at work in any computational work. Computational complexity focuses on the resources needed to compute a problem -- its ``hardness'' --~ rather than if it is computable at all -- its ``computability''. Emerging as a sub-discipline of theoretical computer science only in the late 1960s, its foundational (unsolved) problem known as ``P vs.~NP'' is deeply connected to the philosophical side of artificial intelligence research\footnote{See \autocite{aaronson_2011} for a more thorough exploration of this intersection.}, namely to potential ways of proving the intelligence of machines beyond the confines of the Turing test. \autocite{turing_1950} Here, it serves to argue that, even given an extremely small timeline for search, the ``biological complexity'' of protein folding must be much more efficient than a random search. As such, there should be an acting force that guides amino acids to fold in their particular three-directional configurations, and computation must mirror the resultant more efficient search. Computational efforts thus cannot sample every possible configuration; they should, instead, aim to represent or model the biological process of folding at a similar level of complexity. One of these models of representation finds continuity from the physics-based molecular dynamics to a thermodynamic consideration of proteins, by way of information theory.

\section{The Language Paradigm in Biology}\label{the-language-paradigm-in-biology}

\subsection{From Thermodynamics\ldots{}}\label{from-thermodynamics}

Around the same time as molecular dynamics, biochemist Barry Robson proposed an information theoretical analysis of protein folding in his 1974 paper, ``Analysis of the Code Relating Sequence to Conformation in Globular Proteins''.\footnote{Robson's paper is a precursor to the established GOR method of secondary structure prediction, named after the scientists Jean Garnier, David Osguthorpe, and Robson himself. \autocite{robson_1974}.} As Robson writes: ``The conformational free-energy surface of a protein is large, multi-dimensional and occupied by many minima. Since the deepest minimum is {[}\ldots{}{]} only discovered when it is shown that there is none deeper, a prediction by the theoretical conformational energy approach is well beyond the capabilities of any contemporary computer.'' \autocite[854]{robson_1974} Several points are important to highlight from this short passage. First is the acknowledgment of the computational expense of searching for the global minimum ``free-energy'' path to conformation of a protein. Second is this very framing of the theoretical problem of protein folding as a problem of energy. Proteins are conceived of as a thermodynamic system that follows general, physical principles of energy efficiency and stability. It is this physical view that allows Robson to understand protein folding as an information-theoretic problem. In the thermodynamic configuration, there are a number of microstates that could lead to the free-energy macrostate. \autocite{sharp_2015} This problem of translation between micro and macrostates is the continuous measure between thermodynamics and information theory. It is formalized as entropy, whose formulas, whether physical chemist Josiah Willard Gibbs's or information theorist Claude Shannon's, are formally identical. \autocite{schwartz_2019} Robson thus proposes that a suitable method, marrying statistical and theoretical conformational energy approaches, is to treat the ``sequence of amino acid residues and the sequence residue conformations as two messages, the second of which is derived from the first by a translation process.'' \autocite[854]{robson_1974}

\subsection{\ldots{} to Information Theory \ldots{}}\label{to-information-theory}

Following Shannon, this is a clear scenario where modeling information becomes useful. What is important is not necessarily how the protein folds, but how the protein chooses to fold. As Shannon famously writes: ``The fundamental problem of communication is that of reproducing at one point either exactly or approximately a message selected at another point. Frequently the messages have meaning {[}\ldots{]} these semantic aspects of communication are irrelevant to the engineering problem. The significant aspect is that the actual message is one selected from a set of possible messages.'' \autocite[1]{shannon_1948} In the thermodynamic model of protein folding, the analogy would be mapping the free-energy state that dictates the selection of a conformation path. The thermodynamic framing, as Robson shows, allows for both conceptual and mathematical continuities with information theory as a method of predicting protein folding. 

Thus, from the early days of protein folding research, and decades before machine learning became relevant in structural biology, amino acid sequences were thought of as information, and through this framing, as language-like. Crucially, however, amino acid sequences are only language-like within the information-theoretic view of information that explicitly excludes meaning. Robson's usage of information theory, and the current continuities of the thermodynamic hypothesis,\footnote{It should be noted that this paper is actually a challenge to this prevailing view, but still acknowledges the dominance of theorization of protein folding as a thermodynamically favored process. \autocite[1]{sorokina_2022}} foreshadow our larger critique that highlights the transformer architecture's ``non-linguistic'' treatment of language. Information theory, historically, serves as a bridge between structural biology and structural linguistics -- but it might never have been more than a useful analogy.

\subsection{\ldots{} to Linguistics}\label{to-linguistics}

Finally, it should be noted that the centrality of the language paradigm to the operationalization of biological processes is not exclusive to protein folding. Rather, it is perhaps most prominently represented in the longstanding effort to ``decipher'' the ``Book of Life'' \autocite{kay_2000} -- the genetic code -- by understanding molecular biology through the lens of structural linguistics. Here, Roman Jakobson's contributions to linguistics play an important role. Jakobson, a Russian émigré famous for introducing structuralism to American literary criticism, had a substantial yet lesser-known influence in shaping the discourse of ``DNA linguistics'' \autocites{kay_2000}[see also][]{searls_1992}, exemplified in his early endeavors in identifying a meta-language -- a universal system of symbols applicable to both human languages and natural sign systems. For Jakobson, the universal appeal and symbolic efficiency of communication science were intrinsically tied to his pursuit of this meta-language spanning both cultural and biological domains. Bolstered by the era's aggressive advances in bioengineering, Jakobson confidently asserted that studying DNA was akin to studying languages, given their shared meta-linguistic attributes. \autocite[297-315]{kay_2000} Today, this fundamental entanglement is mirrored, for instance, in cross-disciplinary interests of natural language researchers, who seem to easily move between language processing and challenges like mRNA sequence design optimization for vaccine research. \autocite{zhang_2023}

\section{The Transformer Architecture}\label{the-transformer-architecture}

Given this historical prevalence of the language paradigm in biology, today's ``intuitive'' use of ``language models'' for protein folding connects the early nexus of thermodynamics and information theory, as well as the entanglement of structural biology and structural linguistics, to the transformer architecture.\footnote{It should be noted that we are intentionally skipping over multiple steps in the technical history of structural biology here, including earlier forms of machine learning models developed during the 1980s and 1990s. While a further investigation of these intermediate forms would contribute to the historical study of protein folding, and would likely give additional evidence to the prevalence of the language paradigm in structural biology, our aim here is to show instead how the particular technical innovations introduced by the transformer architecture restructure the language paradigm itself.}

The transformer architecture is a deep learning architecture especially attuned to handling sequential input data. While not generative by design\footnote{Specifying the not-necessarily generative nature of transformers is relevant here, as we will argue that we need to understand ``transformer'' as an epistemic structure separate from, but related to, the epistemic structure of ``natural'' language, and protein ``language'', independent of sequence-completion tasks.}, it rose to prominence as the most innovative aspect of generative architectures like GPT, and, subsequently, pre-trained language models like OpenAI's ChatGPT 4. The literature (both within computer science and critical artificial intelligence studies) is thus ripe with more or less complex descriptions of the transformer architecture's design principles.\footnote{We would like to direct the reader’s attention, in particular to the video tutorial by 3Blue1Brown (https://www.3blue1brown.com/lessons/gpt), the code-based tutorial ``The Annotated Transformer'' (https://nlp.seas.harvard.edu/annotated-transformer/), and Stephen Wolfram’s blog post ``What Is ChatGPT Doing … and Why Does It Work?'' (https://writings.stephenwolfram.com/2023/02/what-is-chatgpt-doing-and-why-does-it-work/).} Thus, we will limit our own summary here to its eponymous core mechanism, the attention mechanism \autocite{vaswani_2017}, which sharply distinguishes the transformer architecture from earlier approaches.

Attention, in simple terms, is a mechanism that directs a model's focus to multiple parts of an input sequence at once. Earlier model architectures based on Markov chains \footnote{Subsequent development of more advanced models, such as RNNs and LSTMs, aimed to address the long-term dependency limitations inherent in Markov chains. However, these models still struggled with long-term dependencies due to issues like the vanishing gradient problem. From a statistical perspective, neither RNNs nor LSTMs exhibit long memory capabilities. See \autocite{li_2022}.} in which only directly preceding tokens would be considered when trying to predict the next token in a sequence ignored, for instance, the impact of tokens near the beginning of a sentence on tokens at the end. That different parts of the input sequence are considered in parallel (mathematically modeled as a matrix multiplication operation) thus not only means that the sequence is processed more effectively, but also that long-range dependencies are taken into account -- which in turn means that more complex input-output mappings are possible. In the case of proteins, if -- following Anfinsen's Dogma -- all information needed to model the folding process is present -- but not immediately visible -- in the amino acid sequence, a sequential input (an amino acid chain) is used to predict a more complex three-dimensional structure (a protein), similarly to how one could use a sentence to predict a forthcoming word or sentence.

However, in this seemingly logical translation from language to biomolecule, a critical question is raised around how exactly computers represent both words and protein structures. Below, we address this issue of the apparent biological-linguistic universality of the transformer architecture by pointing out where the affordances of the model leave the realm of language and enter a separate, autonomous epistemic space which can be adapted to both biological and linguistic phenomena. We pay special attention -- as it were -- to essential but often disregarded pre-processing steps which, counterintuitively, represent the least linguistic aspects of the transformer: subword tokenization, word embedding, and positional encoding.

\subsection{Subword Tokenization}\label{subword-tokenization}

Subword tokenization emerges from a fundamental challenge of natural language processing: there are too many word-level tokens to be efficiently encoded, while character-level encoding is too fine-grained to preserve high-level information. The goal of subword tokenization is to strike a balance between context-rich, yet inefficient word-level tokenization and efficient, yet semantically impoverished character-level tokenization. Subword tokenization breaks up words into smaller, more manageable units (subwords) representing syllables or strings of characters that occur frequently in a textual corpus. For example, the rare word ``refactoring'' can be broken down into recurring units such as ``re,'' ``factor,'' and ``ing.'' These more manageable units still maintain contextual relevance and are semantically salient. At the same time, it is already in this very first pre-processing step that we can see the ``linguistic'' nature of language, its dependence on discrete tokens organized in hierarchical structures, vanish.

Subword tokenization begins with deriving a subword vocabulary from a larger corpus of text. Commonly used algorithms for this derivation include Byte Pair Encoding (BPE), WordPiece, and SentencePiece. By iteratively searching for, and merging character sequences that frequently co-occur, these algorithms establish a vocabulary capable of representing any word in the corpus while maintaining efficient size for tokenization. For instance, in BPE, the algorithm starts with a vocabulary of individual characters and progressively merges the most common sequences of characters into subword units. This process iterates until a desired vocabulary size is reached or the gains in representational efficiency diminish. The resulting vocabulary contains both individual characters (to handle rare words or unique character combinations) and commonly recurring subword units.\footnote{The foundational premise here is that words are morphologically diverse. That is, words can be broken down into more fundamental meaningful constituents, called morphemes, which facilitate flexibility and creativity in language evolution (such as accommodating neologisms).}

However, when we break up words into subwords, we leave the realm of ``words'' as discrete symbols and enter the realm of ``tokens''. While still representing ideas, or parts of ideas, tokens cease to be independent meaning-making units. Rather, they are concepts in and of themselves, thus introducing a layer of artificial continuity into the discrete space of natural language. We use ``continuity'' here to mean the simulated, floating-point continuity of the computer, rather than a truly continuous signal, albeit a deeper philosophical argument could be made about the epistemic potential afforded by both modalities -- but it is important to keep in mind that ontological questions never apply in the realm of computation, which is built on an irreversible process of discretization. This artificial continuity, then, is further amplified by the next pre-processing step of the transformer architecture, word embedding.

\subsection{Word Embedding}\label{word-embedding}

In order for machines to ``read'' and ``understand'' human language, words must be transformed into numbers. This can be achieved through various methods. One such method is one-hot encoding, in which each word in a given input vocabulary of size n is represented as a vector of size n, where the vector contains a 1 at the position corresponding to the word and 0s elsewhere. A one-hot encoded word, however, loses its context entirely.

The engineering attempt to preserve -- again, like in the case of subword tokenization -- ``some of'' the meaning of the input sequence, then, comes in the form of word embedding. Either learned during the training process or sourced from another model, embedding matrices position each word in relation to\footnote{This relational aspect of word embedding has often been compared to poststructuralist notions of meaning, particularly Derrida’s notion of différance. It should be noted, however, that the relational saliency of embedded words is a product only of their operationalization: only words that are, in fact, numbers, gain relational saliency.} other words in the input sequence. For instance, for the popular Word2vec model, if two words frequently appear alongside each other (within a certain window size), then their embedded representations should be likewise spatially proximate. Though this proximity is not arbitrary, it is still constrained by the continuous but finite dimensions of the embedding vector space (typically in the hundreds or even thousands) that may not capture all possible nuances in the relation between words. As such, relationships between words are to be understood as contextually determined and constrained by the fundamental limitations inherent in translating words to numbers. Nevertheless, as the famous analogy test\footnote{Word embedding models can answer simple analogy questions like ``king is to man what woman is to ?''} shows, some semantic aspects are indeed preserved.

The attention mechanism operates on words as represented by these high-dimensional vectors. As the output of the model, before any post-processing, is produced also in the form of high-dimensional vectors, one must ``de-embed'' them. De-embedding seeks to convert vectors back into words. The problem with de-embedding, however, is that the relational projection never yields perfect matches. Instead -- due to the continuous nature of the embedding space -- it results in a range of probable words, necessitating a further step -- sampling\footnote{See Rita Raley’s forthcoming essay on sampling in \autocite{bajohr_2024}.} -- to transform these possibilities into a concrete textual output. The continuous embedding space, in a sense, must be ``collapsed'' to generate discrete words. Words -- or rather, tokens, as word embedding follows subword tokenization in the preprocessing pipeline -- are thus translated into continuous representations before they receive any \textit{attention} (in both the technical and the literal sense). They are treated as spatially contiguous rather than symbolically discrete. This leads to the scenario where certain tokens exist ``in-between'' the tokens of the vocabulary that would best fit the transformer's output vectors. \autocite[see][]{offert_2017} These tokens are computationally existent but virtually non-existent in human language.

\subsection{Positional Encoding}\label{positional-encoding}

Positional encoding is used to capture positional information in token sequences which would otherwise be lost due to the position-insensitive attention mechanism. This encoding process highlights that embedding vectors represent not tokens per se but an overlay of heterogeneous abstract concepts. Unlike Long Short-Term Memory (LSTM) models that process tokens sequentially, attention-based transformers process tokens in parallel, significantly improving computational efficiency. However, this parallel processing sacrifices positional information (the order of words, for instance) that may play a crucial role in contextualizing the overall representation of a sequence.

To redress this loss -- and to augment the limited positional information provided by the embedding process -- positional encoding methods project positional information of tokens into embedding vectors, which are then added to token embeddings themselves to form a complex (in the precise mathematical sense) vector representation. The attention mechanism, thus, operates on the sum of positional and token information. In a more special case, the often-employed BERT model \autocite{devlin_2018} incorporates a ``{[}CLS{]}'' (Classification) token at the beginning of the input sequence during the pre-processing stage. This token serves as an aggregate representation of the entire input sequence, preserving contextualized information. Although it is processed similarly to normal tokens, the {[}CLS{]} token does not carry the same kind of semantic information as other tokens.~

Interestingly, it is somewhat unclear if by linearly adding two distinct sources of information together the vector representation can effectively capture correlations between tokens and their positions. For example, the position of words is crucial for determining the overall meaning of a sentence, but the extent to which such correlation can be captured through positional encoding lacks theoretical support. A recent study suggests that the current positional encoding method introduces a high volume of noise, indicating the method's potential limitations in accurately capturing positional information. \autocite{ke_2020} As a result, the original goal of positional encoding -- to restore positional information lost in parallel processing -- is not fully achieved in practice. Positional encoding struggles to accurately capture contextual information narrowly understood from the perspective of structural linguistics, let alone the performative and worldly dimensions of language in use, which have not yet been formalized through basic computational theories. Regardless of its efficiency, however, positional encoding adds yet another layer of non-linguistic continuity to the pre-processing pipeline.

\section{The ESM2 Model Case Study}\label{the-esm2-model-case-study}

How exactly are the pre-processing techniques described above put to work? What role do they play in the larger ecosystem of ``scientific'' and ``cultural'' models? Here, we need to turn to the actual object of this paper, transformer-based machine learning models of the protein folding process. We focus on one case study in particular -- the ESM2 model produced by Meta \autocite{lin_2023} -- as an example of a protein folding model fully determined by the language paradigm.

One of the most important components of any machine learning system is data\footnote{Multiple large-scale datasets are available for computational protein folding. The oldest one is the Protein Data Bank (PDB), a worldwide repository providing a comprehensive collection of experimentally determined three-dimensional structures of proteins, nucleic acids, and complex assemblies. It is widely used by researchers in the field of structural biology to analyze protein-protein interactions and perform various other kinds of computational analyses of protein structures in addition to protein folding. In addition to the PDB, there are other databases specifically designed to support the study of protein folding, including Protein Folding Database (PFD) (with an emphasis on protein folding kinetics and thermodynamics data) and ProteinNet (designed specifically for deep learning approaches, including diverse protein families and structures).} -- in our case, data that encode those long-range dependencies the transformer architecture aims to decode. The ESM2 model models the MGnify90 dataset \autocite{richardson_2022}, which comprises curated and annotated protein sequences from a wide range of organisms found ``in the wild'', including those found in microbes in the soil, deep in the ocean, and even inside our bodies -- vastly outnumbering those that make up animal and plant life.

Compared to other protein prediction models such as AlphaFold, AlphaFold 2 and RoseTTAFold, the ESM family of models circumvents the need to go through the time-consuming step of multiple sequence alignment.\footnote{The multiple sequence alignment (MSA)-based approach to protein folding, exemplified by models like AlphaFold, AlphaFold 2 and RoseTTAFold, incorporates two transformer blocks to iteratively refine representation generated by the previous step of coevolutionary analysis, where input sequences are compared with protein sequences whose 3D structures were experimentally determined, to search for matches. These components are crucially important when it comes to constructing full-length 3D structures. Thus far, this approach has outperformed the ``language model''-based method, demonstrated by AlphaFold 2’s significant lead at CASP 14 meeting. However, the MSA-based method’s reliance on comparing and matching requires a high computational cost which at the same time limits its effectiveness in predicting those sequences that cannot be identified and matched with homologs. These orphan sequences count for 20\% of all metagenomic sequences. This limitation has paved the way for the ``language model''-based approach, which does not require comparing and matching.} Instead, the ESM architecture requires just the amino acid sequence of the protein as input. The model output is composed of 3-D coordinates that describe the position of each atom in the protein molecule. In that sense, ESM2 is a ``pure'' ``language model'' -- a fact emphasized throughout the research paper and blog post accompanying its development.

Among other methods from natural language processing, ESM2 -- a transformer-based model -- employs ``masked language modelling.'' During its training phase, approximately 15 per cent of the amino acid positions in a sequence are randomly obscured -- either masked, shuffled to a different amino acid, or left unchanged. The model is then trained to predict these obscured residues based on the bi-directional context of sequences, i.e. considering both the preceding and subsequent residues around any given position in a sequence. From this process, the model then learns a ``translation'' -- to speak with Robson and Shannon -- from sequence to structure.  

ESM2, importantly, has been employed to predict three-dimensional protein structures on a large scale, contributing to the so-called \textit{ESM Atlas}, which incorporates predictions from ESM2 and other related models. The \textit{ESM Atlas} provides the ``first large-scale view of the structures of metagenomic proteins encompassing hundreds of millions of proteins.'' \autocite{ESMMeta} As such, it is first and foremost a demonstration of the projected efficiency of transformer-based protein folding. As the project blog post states:  ``Advancements in gene sequencing have made it possible to catalog billions of metagenomic protein sequences. Although we know that these proteins exist, because we have discovered their sequences, understanding their biology is a staggering challenge. Determining the three-dimensional structures for hundreds of millions of proteins experimentally is far beyond the reach of time-intensive laboratory techniques such as X-ray crystallography, which can take weeks to years for a single protein. Computational approaches can give us insight into metagenomics proteins that isn’t possible with experimental techniques.'' \autocite{ESMMeta}

What, then, enables this large-scale success, if we are to believe the authors of the project? A paper investigating the potential of ESM2 to generalize \textit{de novo} proteins beyond existing ones, attributes its success -- perhaps unsurprisingly -- exactly to the model's linguistic nature: ``We hypothesize that there exists a deep underlying grammar in protein sequences that makes it possible for the language model to generalize. {[}\ldots{]} Classically this form generalization has been enabled by an energy function grounded in physics that captures the native folded state. {[}\ldots{]} {[}N{]}ew deep learning approaches may capture something similar to the physical energy. The success of language models on this problem suggests that deep patterns in sequences may offer an alternative path to generalization, independent of an explicit model of the underlying physics.'' \autocite[3]{verkuil_2022}

It is this paradoxical mapping of language and physics -- a mapping that we find already in Robson -- that structures the project's epistemic perspective. Language and physics are different but the same, models of nature but also models of each other. ``Like the text of an essay or letter,'' the blog post reads, ``proteins can be written as sequences of characters. {[}\ldots{]} Learning to read this language of biology poses extraordinary challenges. While a protein sequence and a passage of text can both be written down as characters, there are deep and fundamental differences between them. A protein sequence describes the chemical structure of a molecule, which folds into a complex three-dimensional shape according to the laws of physics.''

In large language models, we witness emergent behaviors\footnote{The degree to which these behaviors are truly emergent is contested, see for instance \autocite{wei_2022} and \autocite{schaeffer_2024}.} as the model scales up. The rules of grammar, for instance, are never explicitly ``taught'' to any large language model, they are only latent in the training data. The model ``learns'' these rules simply because grammatically-correct language has an almost infinitely higher likelihood to appear in the data than nonsense. At the same time, as the popular image of the \textit{Library of Babel}, most famously described by Borges, suggests, the number of meaningful (i.e. syntactically and semantically valid) token sequences is vanishingly small for natural languages. Token sequences structured by rules thus appear in sharp contrast to random ones, which makes learning easier. A parallel form of learning is assumed to take place in transformer-based models trained on protein sequences. The strength of ESM2 is thus premised on the assumption that the generative forces behind meaning-making in human language may have analogues in the biological ``language'' of proteins. What the authors suggest is, almost word for word, albeit in terms of the current generation of ``language technology,'' Robson's leap from thermodynamics to information theory almost half a century prior. Transformers learn the rules of folding akin to a physical model, but ``differently.'' The ``deep grammar'' of symbolized bases, in other words, can represent the epistemic structure of physical reality just as well as its actual epistemic structure (after all, it is physical laws that govern the molecular processes of protein folding.).

\section{Discussion}\label{discussion}

Here, then, lies the epistemic dilemma at the core of contemporary transformer models. 

The operations performed by subword tokenization, word embedding, and positional encoding suggest that whatever transformers operate on and learn from has lost its ``linguistic'' character. We might collect a large datasets of ``natural'' language to train a transformer model but what it actually operates on is anything but ``natural''. Pragmatically, we would not recognize the results of the three preprocessing steps described above as language-like anymore.

At the same time, before preprocessing even takes place we have arguably already entered a ``non-linguistic'' space, even if we bracket the bottom parts of the stack.\footnote{For the sake of our argument there is \textit{only} software, to paraphrase (and invert) Kittler's famous ontological perspective.} As many others have pointed out \autocite{shoemaker_20XX}, basic text encoding and its many idiosyncrasies already dilutes the linguistic character of language before it is fed to any software in the proper sense: language is not a medium of computation.\footnote{Yet. See \autocite{offert_2023a}.} We thus have to be more precise about what exactly is being lost here, and about what exactly we mean by ``linguistic'' or ``language-(un)like'' in the first place.

Crucially, it is exactly the fact that proteins are treated like language that powers protein folding models. Although it is true that amino acids are not letters and amino acid sequences are not sentences it is equally true that they must become letters and sentences in order for them to be inputted as data into a transformer.\footnote{We are quoting reviewer two of the paper here, and would like to thank them for their invaluable input on this argument.} The language paradigm, in other words, is, and always has been, an operational paradigm: not only a way to \textit{read} reality through the lens of language, but a way to \textit{understand}. 

More precisely: while the language paradigm breaks down if we understand it as doing the representational work of ``encoding'', it becomes productive if we understand it as providing a generalized framework for the discrete modeling of continuous systems, exactly in the sense of information theory. It is no coincidence, then, that Hans Blumenberg, in the final chapter of \textit{The Readability of the World}, quotes quantum physicist Erwin Schrödinger stating: ``The great revelation of quantum theory was that features of discreteness were discovered in the Book of Nature, in a context in which anything but continuity seemed to be absurd according to the views held until then.''\footnote{Op. cit. \autocite[p. 312]{blumenberg_2022}.}

But why not move directly to information theory then, why the detour through language, arguably only the case study of information theory. If language, in transformers, is a special case, if transformers are ``general pattern machines'' \autocite{mirchandani_2023}, if the extraction of patterns from linguistic data is just one application, why, then, do we stick to the metaphors derived from the language paradigm even at the fringes of their usefulness -- again, amino acids are not letters and amino acid sequences are not sentences. 

The answer, we would like to suggest, is operationalized intuition. As Peter Galison famously argues in \textit{Image and Logic} \autocite{galison_1997}, and perhaps even more explicitly in ``Images Scatter into Data, Data Gather into Images'': neither ``the ‘pictorial-representative'' nor the ``analytical-logical'' exist as fixed positions in scientific imaging \autocite[p. 322]{galison_2002}. And both are needed, desperately, to learn, to know: ``We must have scientific images because only images can teach us. We are human, and as such, we depend on specificity and materiality to learn. Only pictures can develop within us the intuition needed to proceed further towards abstraction. [...] And yet, we cannot have images because images deceive. Pictures create artifactual expectations, they incline us to reason on false premises. We are human, and as such are easily led astray by the siren call of material specificity. Logic, not imagery, is the acid test of truth that strips away the shoddy inferences that accompany the mis-seeing eye'' \autocite[p. 300]{galison_2002}. Language, we might state then, provides the medium for another, particular kind of intuition that is not afforded by physics -- that is not afforded, in fact, by either ``image'' or ``logic'': an intuition about \textit{sequences}. There lie the roots of the language paradigm as \textit{metaphor} -- and, in the last instance, also as \textit{technique}. It is only in the medium of language that we can reason about sequences, and operations on sequences. Shannon \autocite{shannon_1948} develops his information theory from statistical patterns in the English language, but perhaps more importantly in explicit reference to instances of low and high entropy language in literature, namely Ogden and Joyce \autocite{kittler_2024}. Indeed, the very idea \textit{of} statistical patterns in language can be traced back to literature, namely to Markov's \autocite{markov_2006} analysis of \textit{Eugene Onegin} from which the notion of the Markov model of dependent probabilities emerges.

We are facing, thus, a double transformation, a kind of \textit{epistemic detour}, in protein folding models. To make physical reality amenable to sequential intuition it has to be made linguistic. We just \textit{know} that the secret to protein folding must be expressible as a complex pipeline of sequential operations (and the persistence and success of the language paradigm shows that this intuition has merit). But as we are dealing with unknown (or at least under-formalized) bio-physical processes in protein folding, their transformation into sequential operations is not straightforward -- it is, in fact, precisely non-linear -- and thus left to the machine. That machine -- the transformer -- however, on the one hand cannot process language in the first place, and on the other hand needs continuity (in the computational sense) to model continuity (in the physical sense), which means that our language-reality must immediately be dissolved again, through the preprocessing steps described above. It is only then that enough ``space'' is created, through the introduction of artificial continuity, and quite literally in the gaps between embedded data points, that new knowledge can be produced.\footnote{For a longer and more general version of this argument, see \autocite{offert_2025}.} In other words: in protein folding, we take an epistemic detour through language because we desire the sequence-based reasoning afforded by language without rendering physical reality entirely metaphorical. The transformer, then, does exactly more and less than language. It removes almost all non-operationalizable sense-making dimensions (think, for instance of inter- or paratextuality, of performativity and contingency, or of anything metaphorical that is more complex than a simple analogy) -- but it also adds new sense-making dimensions through subword tokenization, word embedding, and positional encoding.

If we began by suggesting that by all but ignoring the vast landscape of scientific models within and beyond structural biology, critical artificial intelligence studies is missing an entire angle of critique, we can  now specify: it is exactly in the epistemic structure of the transformer architecture that ``cultural'' and ``scientific'' concepts become entangled. Transformers are ``cultural'' because they render physical reality ``linguistic'' to afford sequential reasoning, and they are ``scientific'' because this ``linguistic'' version of physical reality is only created to be immediately resolved back into numbers (or, more precisely, vectors). The persistence of the language paradigm in biology has always suggested a trading zone. In the transformer age, this trading zone has become material. At the same time, no recurrent neural network, no deep convolutional neural network, and no diffusion network affords the kind of knowledge afforded by the transformer. Architecture- and model-specificity, in other words, matters -- and this suggests much work ahead for critical artificial intelligence studies.

\section{Acknowledgments}\label{acknowledgements}

The authors would like to thank Elena Aronova, Anna Schewelew, and Rita Raley, as well as the two anonymous reviewers, for their time and helpful suggestions.

\printbibliography

\end{document}